\documentclass[letterpaper,twocolumn,10pt]{article}
\usepackage{usenix}
\usepackage{cuted} 
\usepackage{tikz}
\usepackage{amsmath}

\usepackage{filecontents}

\usepackage{amsmath,amsfonts}
\usepackage{multirow}
\usepackage{colortbl}
\usepackage[breakable]{tcolorbox}
\usepackage[ruled, vlined, linesnumbered]{algorithm2e}
\usepackage{booktabs}

\begin{document}


\title{RobustMask: Certified Robustness against Adversarial Neural Ranking Attack via Randomized Masking}

\author{
Jiawei Liu\textsuperscript{1},
Zhuo Chen\textsuperscript{1},
Rui Zhu\textsuperscript{2},
Miaokun Chen\textsuperscript{1},
Yuyang Gong\textsuperscript{1},
Wei Lu\textsuperscript{1}\thanks{Corresponding authors.},
XiaoFeng Wang\textsuperscript{3}\footnotemark[1],
\\
\textsuperscript{1}Wuhan University,
\textsuperscript{2}Yale University,
\textsuperscript{3}Nanyang Technological University
}

\maketitle

\begin{abstract}

Neural ranking models have achieved remarkable progress and are now widely deployed in real-world applications such as Retrieval-Augmented Generation (RAG). However, like other neural architectures, they remain vulnerable to adversarial manipulations: subtle character-, word-, or phrase-level perturbations can poison retrieval results and artificially promote targeted candidates, undermining the integrity of search engines and downstream systems. Existing defenses either rely on heuristics with poor generalization or on certified methods that assume overly strong adversarial knowledge, limiting their practical use.
To address these challenges, we propose \textit{RobustMask}, a novel defense that combines the context-prediction capability of pretrained language models with a randomized masking-based smoothing mechanism. Our approach strengthens neural ranking models against adversarial perturbations at the character, word, and phrase levels. Leveraging both the pairwise comparison ability of ranking models and probabilistic statistical analysis, we provide a theoretical proof of RobustMask’s certified top-K robustness. Extensive experiments further demonstrate that RobustMask successfully certifies over 20\% of candidate documents within the top-10 ranking positions against adversarial perturbations affecting up to 30\% of their content. These results highlight the effectiveness of RobustMask in enhancing the adversarial robustness of neural ranking models, marking a significant step toward providing stronger security guarantees for real-world retrieval systems.
\end{abstract}

\section{Introduction}

Neural Ranking Models (NRMs), particularly those built on pre-trained language models, have demonstrated remarkable success in information retrieval. Yet, they remain inherently susceptible to a wide range of adversarial exploits that can significantly degrade performance under malicious attacks, exposing a critical lack of robustness. For example, adversaries can manipulate the behavior of the model through carefully crafted inputs or operations to launch information manipulation attacks~\cite{liu2022order}. Such exploits not only compromise the quality of search results, but also pose serious security risks to retrieval systems and retrieval-augmented LLM applications. As a result, a central challenge in information retrieval lies in strengthening the robustness of NRMs and designing effective defenses against information manipulation attacks.

\vspace{2pt}\noindent\textbf{Robustness in ranking models}. More specifically, the lack of robustness in information retrieval models can lead to degraded user experience, lower customer retention, reduced product competitiveness, and even deliberate exploitation for information manipulation, with the potential to influence public opinion and perception~\cite{liu2022order,rubin2012information,epstein2015search,goren2018ranking}. Adversarial attacks against retrieval models can arise in various scenarios, leveraging tactics such as fake news, misleading advertisements, and biased content to advance objectives in domains ranging from political propaganda and marketing to social media manipulation and malicious search engine optimization~\cite{chen2024research,zhou2024adversarial}.

Prior efforts to improve the robustness of text ranking models have been largely empirical. Some studies attempt to detect adversarial samples using features such as TF-IDF~\cite{zhou2009osd} and perplexity~\cite{song2020adversarial,ma2023abstract}, while others focus on generating adversarial examples based on known attack methods for adversarial training~\cite{chen2023dealing}. However, these empirical defenses remain fundamentally limited, as they cannot fully capture the diverse strategies adversaries may employ, and consequently provide only modest gains in robustness, not to mention any guarantee for the protection.

\begin{figure*}[!t]
\centering
\includegraphics[width=0.65\linewidth]{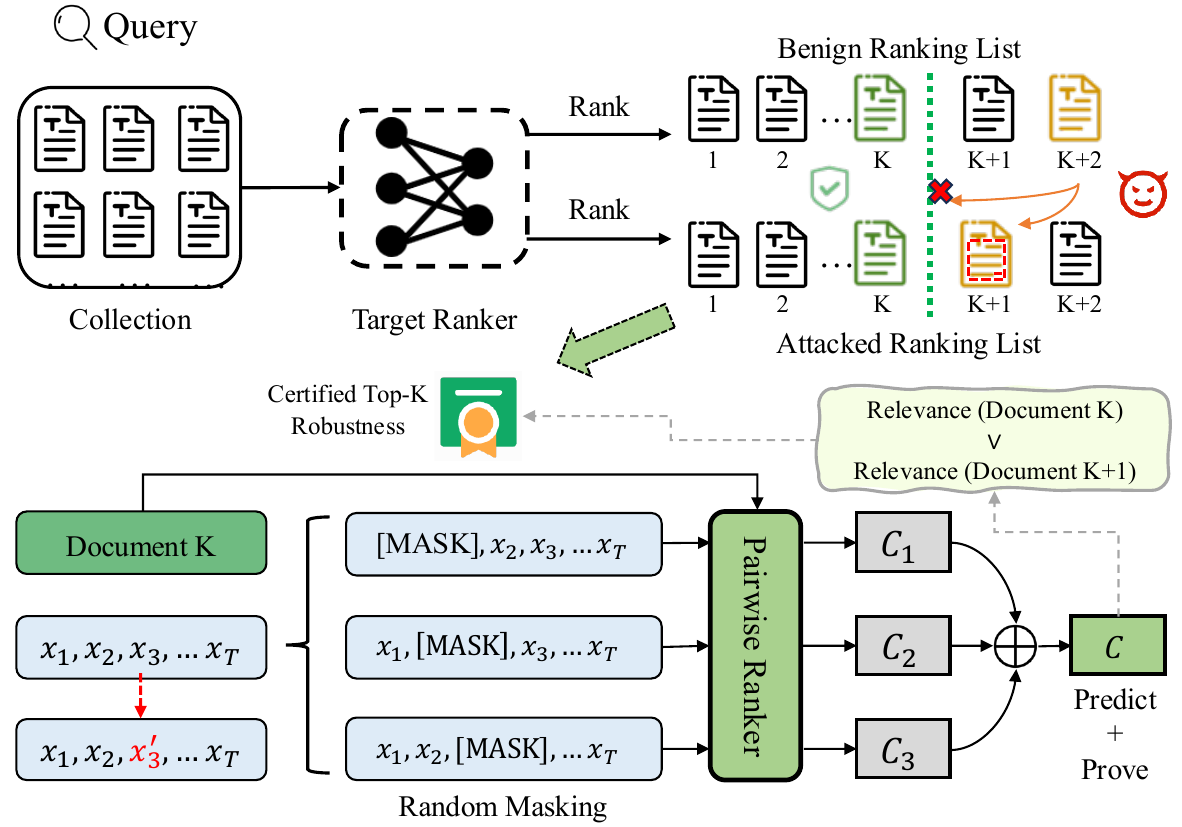}
\caption{The overview of the Certified Top-K Robustness and the architecture of our proposed RobustMask method.}
\label{fig_mpt}
\end{figure*}
To counter unknown or adaptive attacks and break the ongoing arms race between attackers and defenders, it is crucial to establish certifiable robustness guarantees, theoretically assurances that a model’s predictions remain stable under bounded adversarial perturbations. However, progress in this direction remains limited. A notable attempt is the work of Wu et al.~\cite{wu2023prada}, which leverages synonym substitution to generate output ranking scores for randomly perturbed documents, thereby constructing a smoothed ranker based on averaged scores. Using the ranking and statistical properties of these perturbed documents, they derive a criterion for certifying the robustness of the target model. 
However, the key limitation of this approach lies in its strong assumption that the defender has prior knowledge of the adversary’s synonym substitution strategy and lexicon, an unrealistic premise that severely undermines its practical applicability. Moreover, the enhancement comes at the cost of substantial performance degradation on clean data.

\vspace{2pt}\noindent\textbf{Our approach}.  In this paper, we introduce \textit{RobustMask}, a theoretically certifiable robustness enhancement technique for text information retrieval models based on \textit{random masking smoothing}. By leveraging the inherent capabilities of pre-trained models, our randomized masking mechanism constructs a smoothed ranking model that mitigates the impact of diverse adversarial perturbations. Unlike prior approaches, RobustMask is built on more realistic assumptions and provides formal guarantees that a model’s top-K predictions remain stable within a bounded range of adversarial manipulations. 
While it primarily certifies the stability of the top-K retrieved results, which is a limited subset of the candidate list, these prioritized rankings align closely with user-centric information needs. As reflected in standard retrieval metrics such as MRR and NDCG, highly relevant documents provide far greater utility to users than lower-ranked items, whose relevance and impact diminish significantly~\cite{liu2022order,nick2009mean}.

Empirical experiments show that our method substantially outperforms the previous state-of-the-art. Compared with CertDR, which achieves a Top-10 CRQ of 9.5\% \cite{wu2022certified}, we achieve a Top-10 CRQ of 58\%, and we reduce the attack success rate of the PRADA \cite{wu2023prada} from 57.4\% to 16.7\%. Beyond PRADA, a representative word- and character-level adversarial attack, our approach also defends effectively against phrase-level attacks such as keyword stuffing, Collision \cite{song2020adversarial}, and PAT \cite{liu2022order}. RobustMask integrates smoothly with language-model pretraining objectives, thereby minimizing its impact on retrieval efficiency. On MSMARCO dataset, it incurs only a 1\%–2\% drop in MRR@10 and NDCG@10 relative to the original ranking model while delivering superior defense effectiveness, whereas CertDR induces a roughly 3\%–5\% degradation under the same setting. These results demonstrate that RobustMask attains strong adversarial robustness with minimal retrieval performance overhead.
Given the widespread adoption of RAG frameworks in modern LLM applications, our study represents a critical step toward strengthening the adversarial robustness of retrieval systems and safeguarding RAG pipelines against security vulnerabilities.

\vspace{2pt}\noindent\textbf{Contributions}. The contributions of this paper are threefold:

(1) We propose RobustMask, leveraging the mask prediction capability from pre-trained language models and a robust randomized masking smoothing mechanism, to effectively defend neural text ranking models against adversarial perturbations at character, word, and phrase levels.

(2) We theoretically establish the robustness guarantees of RobustMask by using pairwise ranking comparison capabilities and probabilistic statistical methods, providing formal and provable certification for the Top-K ranking robustness of neural ranking models.

(3) We verify the practicality and effectiveness of RobustMask, clearly demonstrating that the method can certify the robustness of over 20\% of candidates in the top-10 rankings against any perturbation up to 30\% of content. It demonstrates significant improvements in robustness over existing methods.

\section{Related works}

\subsection{Neural Text Ranking Models}
Classical text ranking methods primarily rely on exact term matching, typically employing efficient bag-of-words representations such as the BM25 algorithm \cite{robertson1994some}. 
To mitigate the vocabulary mismatch issue inherent in exact term matching, neural ranking models introduced continuous vector representations, such as word2vec \cite{mikolov2013distributed} and Glove \cite{pennington2014glove}, combined with neural network architectures to perform semantic soft-matching. Existing neural ranking frameworks can be broadly categorized into three types: representation-based models \cite{huang2013learning,shen2014latent}, interaction-based models \cite{guo2016deep,hofstatter2020interpretable}, and hybrid models integrating both approaches \cite{mitra2017learning}.
Transformer-based models pretrained with language-modeling objectives, notably BERT \cite{devlin2019bert}, have significantly advanced the state-of-the-art in neural text ranking. \cite{nogueira2019passage} first demonstrated BERT’s effectiveness for text ranking tasks. Subsequently, numerous pretrained transformer-based language models fine-tuned on domain-specific corpora achieved substantial performance gains \cite{dai2019deeper,gao2021coil,pradeep2021expando,craswell2020overview}.
Even in the LLM era, pre-trained models based on BERT architecture are still one of the core solutions for retrieval technology and downstream applications of LLMs, such as RAG \cite{xu2024search,shi2025know,gao2023retrieval}.

\subsection{Adversarial Ranking Attack}

Neural text ranking models have inevitably inherited vulnerabilities characteristic of neural networks, particularly susceptibility to adversarial examples \cite{song2020adversarial,liu2022order}. Such adversarial attacks pose severe security threats by intentionally misleading ranking models into incorrect judgment, while preserving a surface-level fluency that bypasses conventional detection mechanisms. Specifically, adversarial attacks exploit the inherent relational dependencies between queries and ranked candidates, deliberately manipulating textual inputs to deliberately distort rankings.
Typically, adversaries subtly inject semantic perturbations into textual content at varying granularities, including character-based \cite{ebrahimi2018hotflip}, word-level \cite{raval2020one,liu2023black,wu2023prada}, and phrase-level modifications \cite{song2020adversarial,chen2023towards,chen25flip}. Earlier research primarily focused on adversarial attacks under white-box scenarios, which assume full access to model internals and gradients \cite{raval2020one,goren2020ranking,song2020adversarial}. Recent advances have expanded to more practical and realistic black-box scenarios, manipulating semantic relevance to compromise retrieval effectiveness \cite{chen25flip,bigdeli2024empra,gong2025topic}.

\subsection{Defenses against adversarial text ranking attacks}

To mitigate textual adversarial attacks, previous works propose various empirical defense strategies, primarily involving input pre-processing \cite{pruthi2019combating,jones2020robust}, adversarial training \cite{goodfellow2014explaining}, and robust model architectures \cite{miyato2017adversarial}. 
However, empirical defenses inherently face limitations. In practice, they tend to be scenario-specific and often fail when confronted with novel or adaptive adversaries \cite{athalye2018obfuscated}. For instance, defenses effective against character-level perturbations typically lose effectiveness when facing word substitution attacks, and vice versa. Moreover, empirical defense methods generally do not offer rigorous robustness guarantees about the model’s performance within clearly defined perturbation bounds.

Unlike empirical methods, certified robustness defenses seek to provide provable guarantees about model predictions. Such methods certify that, within a certain explicitly defined perturbation set around an input, the model's output remains unchanged, thus delivering theoretically grounded robustness guarantees \cite{zhang2024text,li-etal-2021-searching}. Representative certification methods include Interval Bound Propagation (IBP) \cite{jia2019certified}, linear relaxation \cite{huang2019achieving}, and randomized smoothing \cite{cohen2019certified,ye2020safer,zeng2023certified}. Particularly, randomized smoothing’s model-agnostic nature and scalability led to its notable application in textual robustness settings.

For adversarial attacks specifically targeting text ranking tasks, relatively few studies have been conducted. Chen et al. \cite{chen2023dealing} try enhance the robustness and effectiveness of BERT-based re-ranking models in the presence of textual noise. Additionally, Chen et al. \cite{chen2023defense} investigated defense against adversarial ranking attacks through detection approaches. They evaluated the efficacy of various detection methods, including unsupervised methods relying on spamicity scores, perplexity, and linguistic acceptability measures, as well as supervised classification-based detectors leveraging BERT and RoBERTa representations. 
Regarding certified robustness specifically in adversarial text ranking attacks, Wu et al. \cite{wu2022certified} propose CertDR, a certified defensive approach built upon randomized smoothing techniques designed to enhance the robustness of text ranking models specifically against synonym-based word substitution attacks. Nonetheless, their certification assumptions, limiting perturbations exclusively to synonym substitutions, may appear overly restrictive and unrealistic for many practical scenarios.

\section{Preliminaries}
\subsection{Adversarial Textual Ranking Attack}

In this paper, we primarily focus on the re-ranking phase within the two-stage retrieval process (recall-rerank). For the re-ranking phase, given a textual query $q$ and a set of candidate documents $\{x_1, x_2, \cdots, x_N\}$, the ranking model can calculate the relevance score $s(q, x_i)$ for each candidate document $x_i \in \mathcal{D}$, and subsequently generate a ranked list $L = {x_1, x_2, \cdots, x_N}$, such that $x_1 \succ x_2 \succ \cdots \succ x_N$, if $s(q, x_1) > s(q, x_2) > \cdots > s(q, x_N)$. Here, it is assumed that the relevance probability score $s(q, x_i)$ ranges between 0 and 1, which can be straightforwardly achieved by applying a sigmoid or softmax function to the logits output from the ranking model.

Attacks aimed at manipulating neural text ranking models primarily seek to identify a sequence of text that intentionally disrupts the intended order of ranking. This paper attempts to propose a theoretically grounded method to enhance robustness against such attacks. To maintain generalizability, the focus is on manipulation attacks based on word substitution. For example, assume $s(q, x_i) > s(q, x_j)$, where $ x_j = \{w_1, w_2, \ldots, w_T\}$. An adversarial manipulator might generate an adversarial sample $x_j' = \{w_1', w_2', \ldots, w_T'\}$ based on the query $q$ and the document $ x_j $, by perturbing up to $R \leq T$ words in $x_j$, in such a way that the target ranking model makes an error. For an adversarial information manipulation attacker, an adversarial sample $x_j'$  is considered effective if the following conditions are satisfied:
\begin{align}
    s(q, x_j) < s(q, x_j'), \quad || x_j' - x_j ||_0 \leq R
\end{align}
where $|| x_j' - x_j ||_0 = {\sum}_{t=1}^T I\{w_t \neq w_t'\}$ represents the Hamming distance, and $\mathbb{I}\{\cdot\}$ is an indicator function. Here,  $S_x := \{x' : || x_j' - x_j ||_0 \leq R\}$ denotes the set of candidate adversarial samples available to the attacker.
Ideally, all $x_j' \in S_x$ would possess the same meaning from the perspective of human readers, yet could lead to different relevance judgments by the model, potentially elevating their rank according to the target ranking model. For character-level perturbations in English, $w_t'$ might be a visually or phonetically similar typographical or spelling error replacement for $w_t$. For word-level perturbations, $w_t'$ could be a synonym of $w_t$ selected based on the attacker's adversarial strategy. In terms of phrase-level, it is formed by multiple combinations of character-level and word-level perturbations. Practically, the defender does not have prior knowledge of these substitution attack strategies.

\subsection{Certified Top-K Robustness}
Inspired by the certified Top-K robustness definition for ranking models introduced by Wu et al. \cite{wu2022certified}, and the definition of certified robustness for text classification models provided by Ye et al. \cite{ye2020safer}, a ranking model can be considered certified robust if it guarantees that adversarial manipulations of the input will always result in the failure of the attack. This means that regardless of how the attacker modifies the input, the ranking model consistently maintains its effectiveness and integrity of Top-K results, thereby thwarting any attempt to disrupt the intended order of results.

It is well known that in practical search scenarios, users are more concerned with the top-ranked results rather than the lower-ranked candidates, as continued exploration of search results often leads to a sharp decline in click-through rates and traffic. Widely-used retrieval ranking evaluation metrics, such as Recall@N (R@n), normalized Discounted Cumulative Gain (nDCG), and Mean Reciprocal Rank (MRR), also emphasize the importance of top-ranked candidates. Therefore, protecting the top-ranked relevant candidates is crucial not only for downstream applications but also for ensuring the reliability of widely-used ranking metrics \cite{wu2022certified,xia2009statistical}.
A ranking model $s$ can be considered certified Top-K robust if it can ensure that no document ranked beyond the top $K$ can be promoted into the top $K$ positions of the ranking list $L$ through adversarial attacks. This implies that the ranking model $s$ is robust against adversarial manipulations that attempt to alter the ranking of documents within the critical top $K$ positions.

Based on this, the certified Top-K robustness in the context of information retrieval can be formally defined as follows:
Given a query $q$ and the corresponding ranking list $L_q$ produced by a ranking model $s$, we suppose that an attacker conducts adversarial word substitution attacks with specific intensity on any document $d \in L_q[K+1:]$. If the ranking model $s$ can ensure that these attacked documents $x' \in S_x$ remain outside the top $K$ positions of the ranking list, then the ranking model $s$ is considered certified Top-K robust against word substitution attacks, i.e.:
\begin{align}
    Rank(s(q,x'))>K,\ \forall x \in L_q[K+1:] \ \&\ \forall x' \in S_x
\end{align}
where $\text{Rank}(s(q, x'))$ represents the ranking position of the adversarially modified document $x'$ in the ranking list $L_q$ given by the ranking model $s$.

\section{RobustMask}
\subsection{Method Overview}
The core motivation behind the proposed RobustMask method is as follows: if a sufficient number of words are randomly masked from a text, and relatively few words are deliberately perturbed, it becomes unlikely that all perturbed words, which are selected through adversarial attacks, will appear in the masked text. Retaining only a subset of these malicious words is usually insufficient to deceive the text classifier. Thus, to avoid the computationally expensive combinatorial optimization problem, e.g., enumerating all candidate adversarial documents in $S_x$. Inspired by the concept of random smoothing, this paper proposes a defensive method, RobustMask, to enhance the robustness of the baseline text ranking model $s$.

Considering that pre-trained models utilize a masked language modeling (MLM) task during pre-training, leveraging their capability to predict masked tokens can enhance the model's contextual understanding and robustness \cite{devlin2019bert,liu2019roberta}. In this paper, we introduce this masking mechanism to improve the certified robustness of ranking models. This approach constructs a smoothed ranking model as a substitute for the target model, thereby avoiding the exponential computational cost of certified verification.
In conjunction with the relative position relationships in ranking lists, this paper transforms the pointwise relevance score prediction task into a pairwise relative relevance prediction task. This is achieved by repeatedly performing random masking operations on the input text to generate numerous masked copies of the text. The base pairwise ranking model is then used to classify each masked text. The final robust relative relevance judgment is made using a ``voting'' method.

Many studies have shown \cite{xiang2021patchguard,xiang2021patchguard++,huang2023patchcensor,zeng2023certified} that these adversarial examples themselves are highly sensitive and fragile, easily affected by small random perturbations. If some words are randomly masked before feeding the adversarial examples into the classifier, it is more likely that the erroneous predictions of the adversarial examples will be corrected to the correct predictions.

\subsection{Theoretical Certification Robustness}
In this paper, we adopt $\mathbf{x} \ominus \mathbf{x}'$ to denote the set of token coordinate indices where $\mathbf{x}$ and $\mathbf{x}'$ differ. Consequently, we have $|\mathbf{x} \ominus \mathbf{x}'| = ||\mathbf{x} - \mathbf{x}'||_0$. For example, suppose $\mathbf{x} = \text{``A B C D E F''}$ and $\mathbf{x}' = \text{``A B G H E J''}$. Then we have $\mathbf{x} \ominus \mathbf{x}' = \{3, 4, 6\}$ and $|\mathbf{x} \ominus \mathbf{x}'| = 3$. Additionally, let $\mathcal{D} = \{1, 2, \ldots, T\}$ be the set of coordinate indices. We denote $\mathcal{I}(T, k) \subseteq \mathcal{P}(\mathcal{D})$ as the set of all possible subsets of $\mathcal{D}$ containing $k$ unique coordinate indices, where $\mathcal{P}(\mathcal{D})$ is the power set of $\mathcal{D}$. $\mathcal{U}(T, k)$ represents the uniform distribution over $\mathcal{I}(T, k)$. Sampling from $\mathcal{U}(T, k)$ means uniformly sampling $k$ coordinate indices from the set of $T$ coordinate indices without replacement. For example, a possible element $\mathcal{H}$ sampled from $\mathcal{U}(6, 3)$ could be $\{1,2,5\}$.

Define a masking operation $\mathcal{M}: \mathbf{x} \times \mathcal{I}(T, k) \rightarrow \mathbf{x}_{\text{mask}}$, where $\mathbf{x}_{\text{mask}}$ is a text with some of its words masked. This operation takes as input a text $\mathbf{x}$ of length $L$ and a set of coordinate indices, and outputs the masked text where all words except those at the specified coordinate indices are replaced with a special token [MASK]. Since the [MASK] token is used by pre-trained language models during their pre-training phase \cite{devlin2019bert,liu2019roberta}, we will also use this token in this paper to facilitate the efficient utilization of the pre-trained knowledge. For example, $\mathcal{M}(\text{``A B G H E J''}, {1, 2, 5}) = \text{``A B [MASK] [MASK] E [MASK]''}$.

Previous work \cite{wu2022certified} directly utilized pointwise scores to determine relative relevance, employing a predefined perturbation vocabulary for a two-stage proof based on a bi-level min-max optimization. This paper eliminates the dependency on a predefined vocabulary, allowing adversarial word replacement attacks to use any substitute words. 

Specifically, given a query $q$ and the corresponding ranking list $L_q$ provided by the ranking model $s$, for a document $x \in L_q[K+1:]$, the model $s$ is Top-K robust to an adversarial document sample $x'$ if it satisfies $s(q, x_K) > s(q, x')$. Taking the relative relevance order into consideration, it is only necessary to prove that $s(q, x_K) > s(q, x_{K+1}')$, where $x_{K+1}'$ is the adversarial sample corresponding to the document ranked at position $K+1$.
To ensure the ranking model performs consistently and robustly on the data subject to masking operations, we fine-tuned the model. We let the model $f$ be a pairwise relative relevance judgment classifier that performs the mapping $f: \mathcal{X}_{\text{mask}} \rightarrow \mathcal{Y} $. The function $f$ is trained to classify the triplet $($ query $q$, document 1, document 2 $)$ based on the relative relevance between document 1 and document 2, where one of the documents is $x_K$ and the other is $x_{K+1}$. Specifically, the input can be represented as either $[q, x_K, x_{K+1}]$ or $[q, x_{K+1}, x_K]$, with the label set $\mathcal{Y} = {0, 1}$.
Since the concatenation method being one of the most effective approaches for achieving superior re-ranking performance, we also employ the concatenation method as the fundamental structure of the model. Formally, the query $q$ and a pair of candidate items $(x_i, x_j)$ are concatenated using the [SEP] and [CLS] tokens. The relevance score is calculated through a linear layer  $W \in \mathbb{R}^{768 \times 2}$:
\renewcommand{\arraystretch}{1.5}
\begin{align}
    \mathbf{s}_i = \text{LM}([\text{CLS};q;\text{SEP};x_i;\text{SEP}])*W \\
    \mathbf{s}_j = \text{LM}([\text{CLS};q;\text{SEP};x_j;\text{SEP}])*W
\end{align}
where $\mathbf{s}_i$ and $\mathbf{s}_j \in \mathbb{R}^2$ represent the relevance scorers for the positive and negative examples, respectively (also denoted as $s_{\text{pos}}$ and $s_{\text{neg}}$). $LM(\cdot)$ denotes the representational embedding vector derived by the model $LM$ for a given text input. In practice, the model $LM$ used in this study is a BERT model with shared parameters, which is trained on the two concatenated inputs. The loss is calculated as following:
\begin{align}
    \mathcal{L}(q,x_i,x_j) = - y_{i,j} log(softmax(\mathbf{s}_i - \mathbf{s}_j)
\end{align}
where $y_{(i,j)} \in \mathcal{Y}$ represents the one-hot label for $[q, x_i, x_j]$. To ensure a balanced distribution of labels in the training data, this study generates triplets with a label of 0 by swapping the positions of the candidate with relative positive relevance and the candidate with relative negative relevance from the original data. As a result, the $\overline{\mathrm{g}(\mathbf{x})}$ for aggregated copies classification can be defined as:
\begin{align}
    \overline{\mathrm{g}(\mathbf{x})}=\text{argmax}_{c\in\mathcal{Y}} \underbrace{[\mathbb{P}_{\mathcal{H}\sim \mathcal{U}(T,k)}(f(\mathcal{M}(\mathbf{x},\mathcal{H})=c))]}_{p_c(\mathbf{x})}
\end{align}
where $T$ represents the length of $\mathbf{x}$, and $k$ is the number of unmasked words retained in $\mathbf{x}$, calculated as  $[T - \rho \times T]$. $\rho$ denotes the ratio of masked words. $p_c(\mathbf{x})$ refers to the probability that $f$ returns class $c$ after random masking. The predictions of the smoothed classifier can be shown to be consistent with the input perturbations.

The model for ranking in $\mathrm{g}(\mathbf{x})$ can be defined as:
\begin{align}
    \mathrm{g}(\mathbf{x})=\mathbb{E}_{\mathcal{H} \sim \mathcal{U}(T,k)}[s(q, \mathcal{M}(\boldsymbol{\mathbf{x}}, \mathcal{H}))]
\end{align}

Here, we give the definition of Theorem 1.

\textbf{Theorem 1}: 
Given two triplets $\mathbf{x}$ and $\mathbf{x}'$, where $||\mathbf{\mathbf{x}} - \mathbf{x}'||_0 \leq R$, we can derive:
\begin{align}
    \mathrm{g}(\mathbf{x}') - \mathrm{g}(\mathbf{x}) \leq \alpha \cdot \beta \cdot \Delta
\end{align}
wherein:
\begin{align}
\begin{array}{c}
\alpha=\frac{1}{\binom{T}{k}} \\
\beta = f_{avg}(q, \mathcal{M}(\boldsymbol{\mathbf{x}'}, \mathcal{H})) \\
\Delta=\mathbf{1}-\frac{\binom{T-R}{k}}{\binom{T}{k}}
\end{array}
\end{align}
where $\alpha$ represents randomly sampling positions of any $ k $ words from a text of length $T$ for the purpose of a masking operation. 
$\beta$ represents the average relevance score that the ranker $f$ gives to the masked text $\mathbf{x'}$.
$\Delta$ represents the proportion of all possible masking combinations that are excluded when considering the differences between texts $\mathbf{x}$ and $\mathbf{x'}$. It denotes the overall fraction of masking combinations that are affected by these differences.

\textbf{Proof of Theorem 1}:
Given $\mathcal{H} \sim \mathcal{U}(T, k)$, we can obtain:
\begin{align}
\begin{array}{c}
    \mathrm{g}(\mathbf{x}) = \mathrm{g}(q, \mathcal{M}(\boldsymbol{\mathbf{x}}, \mathcal{H})) \\
    \mathrm{g}(\mathbf{x}') = \mathrm{g}(q, \mathcal{M}(\boldsymbol{\mathbf{x}'}, \mathcal{H}))
\end{array}
\end{align}

By subtracting $\mathrm{g}(\mathbf{x})$ from $\mathrm{g}(\mathbf{x}')$, we obtain:
\begin{align}
\label{main_eq}
\begin{array}{l}
    \quad \mathrm{g}(q, \mathcal{M}(\boldsymbol{\mathbf{x}'}, \mathcal{H})) - \mathrm{g}(q, \mathcal{M}(\boldsymbol{\mathbf{x}}, \mathcal{H})) \\
    = \mathbb{E}[s(q, \mathcal{M}(\boldsymbol{\mathbf{x}'}, \mathcal{H}))] - \mathbb{E}[s(q, \mathcal{M}(\boldsymbol{\mathbf{x}}, \mathcal{H}))] \\
    = \int \mathbb{P}(\mathcal{H}) s(q, \mathcal{M}(\boldsymbol{\mathbf{x}'}, \mathcal{H})) - \int \mathbb{P}(\mathcal{H}) s(q, \mathcal{M}(\boldsymbol{\mathbf{x}}, \mathcal{H})) \\
    = \int \mathbb{P}(\mathcal{H}) [s(q, \mathcal{M}(\boldsymbol{\mathbf{x}'}, \mathcal{H})) -  s(q, \mathcal{M}(\boldsymbol{\mathbf{x}}, \mathcal{H}))]
\end{array}
\end{align}
where $\mathbb{P}(\mathcal{H})$ represents the distribution of $\mathcal{H}$.

In considering whether there is overlap between the two index sets $\mathbf{x} \ominus \mathbf{x}'$ and $\mathcal{H}$, we can transform the expression into: 
\begin{align}
\begin{array}{l}
     \quad \int \mathbb{P}(\mathcal{H}) s(q, \mathcal{M}(\boldsymbol{\mathbf{x}'}, \mathcal{H})) \\
     = \int_{}^{\mathcal{H} \in \{\mathcal{H} | \mathcal{H} \cap (\mathbf{x} \ominus \mathbf{x}') \ne \emptyset \}} \mathbb{P}(\mathcal{H}) s(q, \mathcal{M}(\boldsymbol{\mathbf{x}'}, \mathcal{H})) +  \\
     \quad \int_{}^{\mathcal{H} \in \{\mathcal{H} | \mathcal{H} \cap (\mathbf{x} \ominus \mathbf{x}') = \emptyset \}} \mathbb{P}(\mathcal{H}) s(q, \mathcal{M}(\boldsymbol{\mathbf{x}'}, \mathcal{H}))
\end{array}
\end{align}

It is noteworthy that when there is no overlap between the two index sets $\mathbf{x} \ominus \mathbf{x}'$ and $\mathcal{H}$, this indicates that the mask operation has obscured all the words attacked by the attacker, $\mathbf{x}$ and $\mathbf{x}'$ result in identical text after the mask operation. Therefore, we have the following:
\begin{align}
\begin{array}{l}
     \quad \int_{}^{\mathcal{H} \in \{\mathcal{H} | \mathcal{H} \cap (\mathbf{x} \ominus \mathbf{x}') = \emptyset \}} \mathbb{P}(\mathcal{H}) s(q, \mathcal{M}(\boldsymbol{\mathbf{x}'}, \mathcal{H}))  \\
     = \int_{}^{\mathcal{H} \in \{\mathcal{H} | \mathcal{H} \cap (\mathbf{x} \ominus \mathbf{x}') = \emptyset \}} \mathbb{P}(\mathcal{H}) s(q, \mathcal{M}(\boldsymbol{\mathbf{x}}, \mathcal{H})) \\
\end{array}
\end{align}

Back into the main equation \ref{main_eq}, we have:
\begin{align}
\begin{array}{l}
     \quad \int \mathbb{P}(\mathcal{H}) [s(q, \mathcal{M}(\boldsymbol{\mathbf{x}'}, \mathcal{H})) -  s(q, \mathcal{M}(\boldsymbol{\mathbf{x}}, \mathcal{H}))] \\
     = \int_{}^{\mathcal{H} \in \{\mathcal{H} | \mathcal{H} \cap (\mathbf{x} \ominus \mathbf{x}') \ne \emptyset \}} \mathbb{P}(\mathcal{H}) s(q, \mathcal{M}(\boldsymbol{\mathbf{x}'}, \mathcal{H})) - \\
     \quad \int_{}^{\mathcal{H} \in \{\mathcal{H} | \mathcal{H} \cap (\mathbf{x} \ominus \mathbf{x}') \ne \emptyset \}} \mathbb{P}(\mathcal{H}) s(q, \mathcal{M}(\boldsymbol{\mathbf{x}}, \mathcal{H})) \\
     \leq \int_{}^{\mathcal{H} \in \{\mathcal{H} | \mathcal{H} \cap (\mathbf{x} \ominus \mathbf{x}') \ne \emptyset \}} \mathbb{P}(\mathcal{H}) s(q, \mathcal{M}(\boldsymbol{\mathbf{x}'}, \mathcal{H}))
\end{array}
\end{align}

In this context, $\{\mathcal{H} | \mathcal{H} \cap (\mathbf{x} \ominus \mathbf{x}') \ne \emptyset \}$ signifies that the intersection of $\mathbf{x} \ominus \mathbf{x}'$ and $\mathcal{H}$ is not an empty set, implying that the masking operation has not completely obscured the words attacked by the attacker. We denote the probability of the intersection of $\mathbf{x} \ominus \mathbf{x}'$ and $\mathcal{H}$ being non-empty by $\Delta$:

\begin{align*}
    \Delta &= \mathbb{P}(\mathcal{H} \cap (\mathbf{x} \ominus \mathbf{x}') \ne \emptyset) 
    = \frac{\binom{T-|\mathbf{x} \ominus \mathbf{x}'|}{k}}{\binom{T}{k}} = \frac{\binom{T-||\mathbf{x} - \mathbf{x}'||_0}{k}}{\binom{T}{k}}
\end{align*}
where $\mathbb{P}(\mathcal{H})$ denotes the probability distribution of the mask:
\begin{align*}
\mathbb{P}(\mathcal{H}) &= \mathbb{P}(\mathcal{M}(\boldsymbol{\mathbf{x}}, \mathcal{H})) 
     = \frac{1}{\binom{T}{k}} = \alpha
\end{align*}

Further reasoning is as follows:
\begin{align}
\begin{array}{l}
     \quad \mathrm{g}(q, \mathcal{M}(\boldsymbol{\mathbf{x}'}, \mathcal{H})) - \mathrm{g}(q, \mathcal{M}(\boldsymbol{\mathbf{x}}, \mathcal{H})) \\
     = \int \mathbb{P}(\mathcal{H}) [s(q, \mathcal{M}(\boldsymbol{\mathbf{x}'}, \mathcal{H})) -  s(q, \mathcal{M}(\boldsymbol{\mathbf{x}}, \mathcal{H}))] \\
     = \int_{}^{\mathcal{H} \in \{\mathcal{H} | \mathcal{H} \cap (\mathbf{x} \ominus \mathbf{x}') \ne \emptyset \}} \mathbb{P}(\mathcal{H}) s(q, \mathcal{M}(\boldsymbol{\mathbf{x}'}, \mathcal{H})) - \\
     \quad \int_{}^{\mathcal{H} \in \{\mathcal{H} | \mathcal{H} \cap (\mathbf{x} \ominus \mathbf{x}') \ne \emptyset \}} \mathbb{P}(\mathcal{H}) s(q, \mathcal{M}(\boldsymbol{\mathbf{x}}, \mathcal{H})) \\
     \leq \int_{}^{\mathcal{H} \in \{\mathcal{H} | \mathcal{H} \cap (\mathbf{x} \ominus \mathbf{x}') \ne \emptyset \}} \mathbb{P}(\mathcal{H}) s(q, \mathcal{M}(\boldsymbol{\mathbf{x}'}, \mathcal{H})) \\
     \leq \mathbb{P}(\mathcal{H} \cap (\mathbf{x} \ominus \mathbf{x}') \ne \emptyset) \cdot \mathbb{P}(\mathcal{H}) \cdot f_{avg}(q, \mathcal{M}(\boldsymbol{\mathbf{x}'}, \mathcal{H})) \\
     = \alpha \cdot \beta \cdot \Delta
\end{array}
\end{align}

\textbf{Proposition 1.1}: Given a ranked list $L_{q}$ with respect to a query $q$ and the candidate set of adversarial documents as $S_x$, according to Theorem 1, if
\begin{align}
\begin{array}{c}
     \mathrm{g}(q, \mathcal{M}(\mathbf{x}_k, \mathcal{H})) - \mathrm{g}(q, \mathcal{M}(\mathbf{x}_{k+1}, \mathcal{H})) - \alpha \cdot \beta \cdot \Delta \geq 0
\end{array}
\end{align}

We can certify that the ranking of $\mathbf{x'}$ is lower than that of the k-th document , for all $x \in L_{q}[K+1:]$ and any $\mathbf{x}' \in S_x$:
\begin{align}
\begin{array}{c}
     \max\limits_{x \in L_{q}[K+1:]} \max\limits_{x' \in S_x} \mathrm{g}(q, \mathcal{M}(x', \mathcal{H})) \leq \mathrm{g}(q, \mathcal{M}(x_k, \mathcal{H})) 
\end{array}
\end{align}

\textbf{Proof of Proposition 1.1}:
We use $\delta_L$ to denote the difference between the relevance score of the K-th document in the ranked list $L_{q}$ and the maximum possible relevance score of the document subject to word replacement attack. According to Theorem 1, it can be inferred that:
\begin{align*}
    \delta_L &=  \mathrm{g}(q, \mathcal{M}(x_k, \mathcal{H})) - \max\limits_{x \in L_{q}[K+1:]} \max\limits_{x' \in S_x} \mathrm{g}(q, \mathcal{M}(x', \mathcal{H})) 
\end{align*}

Based on Eq 8, we have:
\begin{align*}
    \mathrm{g}(q, \mathcal{M}(x', \mathcal{H})) \leq \mathrm{g}(q, \mathcal{M}(x, \mathcal{H})) + \alpha \cdot \beta \cdot \Delta
\end{align*}

Accordingly, we can transform $\delta_L$ into:
\begin{align*}
    \delta_L &\geq \mathrm{g}(q, \mathcal{M}(x_k, \mathcal{H})) - \max\limits_{x \in L_{q}[K+1:]} \max\limits_{x' \in S_d} [\mathrm{g}(q, \mathcal{M}(x, \mathcal{H})) + \alpha \cdot \beta \cdot \Delta] \\
    &= \mathrm{g}(q, \mathcal{M}(x_k, \mathcal{H})) - \max\limits_{x \in L_{q}[K+1:]} [\mathrm{g}(q, \mathcal{M}(x, \mathcal{H})) + \alpha \cdot \beta \cdot \Delta] \\
    &= \mathrm{g}(q, \mathcal{M}(x_k, \mathcal{H})) - [\mathrm{g}(q, \mathcal{M}(x_{k+1}, \mathcal{H})) + \alpha \cdot \beta \cdot \Delta] \\
    &= \mathrm{g}(q, \mathcal{M}(x_k, \mathcal{H})) - \mathrm{g}(q, \mathcal{M}(x_{k+1}, \mathcal{H})) - \alpha \cdot \beta \cdot \Delta \\
\end{align*}

So if
\begin{align}
\begin{array}{c}
     \mathrm{g}(q, \mathcal{M}(\mathbf{x}', \mathcal{H})) - \mathrm{g}(q, \mathcal{M}(\mathbf{x}_k, \mathcal{H})) - \alpha \cdot \beta \cdot \Delta > 0
\end{array}
\end{align}

Then for for all $x \in L_{q}[K+1:]$ and any $\mathbf{x}' \in S_x$,we have:
\begin{align*}
    \delta_L &=  \mathrm{g}(q, \mathcal{M}(\boldsymbol{\mathbf{x}_k}, \mathcal{H})) - \max\limits_{x \in L_{q}[K+1:]} \max\limits_{\mathbf{x}' \in S_x} \mathrm{g}(q, \mathcal{M}(\mathbf{x}', \mathcal{H})) \geq 0
\end{align*}
i.e., the ranking of $\mathbf{x}'$ is always lower than that of the k-th document.

Since the mask space can be extremely large, we are unable to access the relevance score $\beta = f_{avg}(q, \mathcal{M}(\boldsymbol{\mathbf{x}'}, \mathcal{H}))$. Therefore, we employ the Monte Carlo estimation method to estimate $\beta$, aiming to approximate it as closely as possible to the true value. We estimate $f_{avg}(q, \mathcal{M}(\boldsymbol{\mathbf{x}'}, \mathcal{H}))$ and $\mathrm{g}(q, \mathcal{M}(x, \mathcal{H}))$ using the following approach:
\begin{align*}
    f_{avg}(q, \mathcal{M}(x', \mathcal{H})) &= \mathbb{E}_{\mathcal{H} \sim \mathcal{U}(T,k)}[f_(q, \mathcal{M}(x', \mathcal{H}))] \\
    &\approx \dfrac{1}{n} \sum_{i=1}^{n} f_(q, \mathcal{M}(x'_{(i)}, \mathcal{H})) \\ 
    \\
    \mathrm{g}(q, \mathcal{M}(x', \mathcal{H})) &= \mathbb{E}_{\mathcal{H} \sim \mathcal{U}(T,k)}[g_(q, \mathcal{M}(x', \mathcal{H}))] \\
    &\approx \dfrac{1}{n} \sum_{i=1}^{n} g_(q, \mathcal{M}(x'_{(i)}, \mathcal{H})) 
\end{align*}

According to the definition of $\beta$, its value must lie between 0 and 1 (being positive and less than 1). Consequently, by inferring the inequality conditions for Proposition 1.1, our approach can generate more stringent provable constraints compared to the method employed by Levine and Feizi (2019), who directly set $\beta$ to 1. It is worth noting that the work by Levine and Feizi was focusing on image classification tasks, assuming that all inputs are of equal length and width (i.e., with a fixed number of pixels). However, for text, it is essential to account for the issue of variable length. Therefore, in establishing the provable robustness for text ranking models as defined in this paper, we define the ranking model, the smoothed ranking model, and the values of $\beta$ and $\Delta$ based on the masking rate $\rho$ (i.e., the proportion of words that can be masked), rather than fixing the number of perturbation units as in previous works \cite{levine2019robustnesscertificatessparseadversarial,wu2022certified}. In practical scenarios, given \( x \) and the perturbation number \( R \), one can attempt different masking rates \( \rho \) to calculate the corresponding values of \( \mathrm{g}(q, \mathcal{M}(x, \mathcal{H})) \), \( \beta \), and \( \Delta \). We can use $r_{radius}$ to denote the Certified Robustness Radius (CRR) on the sample \( x \):
\begin{align*}
    r_{radius} = \max_{\mathrm{g}(x) - \mathrm{g}(x') - \alpha \cdot \beta \cdot \Delta \geq 0} R/T
\end{align*}

\begin{algorithm}[tp] 
        \newcommand\Parameter{\textbf{Parameter: }}
        \newcommand\Functions{\textbf{Functions: }}
        \newcommand\Instructions{\textbf{Instructions: }}
        \caption{The Estimation of $\beta$}  
        \label{alg:algorithm1}  
        \LinesNotNumbered
        \KwIn{query $ \boldsymbol{q}$, document $\boldsymbol{x}$ }
        
        \Parameter{Input text length $T$, Number of reserved words $k$, Number of differential words $r$, Number of perturbed samples $n_r$, Number of masked samples $n_k$, randomly smoothed ranker $g$}\\
        \LinesNumbered
        \KwOut{Estimated  $\beta$}
        
        \SetKwProg{Proc}{Function}{}{}
        \Proc{\textit{BetaEstimator}($x, T, k, r, n_r, n_k, g$): }{
            INIT: $\beta\_list \gets 0$ \\
            $\mathcal{A} \gets sample \: n_r \: elements \: from \: \mathcal{U}(T, r)$ \\
            \For{each $a \in \mathcal{A}$}{
             $\mathcal{B} \gets sample \: n_k \: elements \: from \: \mathcal{U}(T, k)$ \\
                \For{each $b \in \mathcal{B}$}{
                \If{$a \cap b = \emptyset$}{$\mathcal{B}.delete(b)$} 
                $s_g \gets based \: on \: ranker \: g \: and \: set \: B$ \\
                $\beta\_list.add(s_g)$ \\
             }
        }
        $\beta \gets average(\beta\_list)$\\
        \Return{\( \beta \)}\
        }
        \label{algorithm1}
\end{algorithm}

\subsection{Estimation and Validation of $\beta$}
The preceding text has defined and explained $\beta$, and here we will discuss how to estimate it. Since $\beta$ represents the average relevance score that the ranker $g$ gives to the masked text where there are intersections between differences $x$ and $x'$, we can utilize the Monte Carlo method to sample a large number of elements from $U(T, k)$ to estimate the value of $\beta$. To simplify the notation, we denote the value of $|\mathbf{x} \ominus \mathbf{x}'|$ as $r$.

The algorithm, as described in Algorithm \ref{alg:algorithm1}, begins by sampling $n_r$ elements from the set $U(T, r)$. Each element, denoted as $a$, is a set of coordinate indices indicating that the corresponding positions of words have been perturbed by the attacker. For each $a$, $n_k$ elements are further sampled from $U(T, r)$. Each of these elements, denoted as $b$, is also a set of coordinate indices, but represents positions where words have not been masked. Subsequently, elements for which the intersection of $a$ and $b$ is an empty set are removed. Using the remaining elements and the ranker $g$, it is possible to approximate the computation of $\beta$.

\begin{figure}[!t]
    \centering
    \includegraphics[width=1\linewidth]{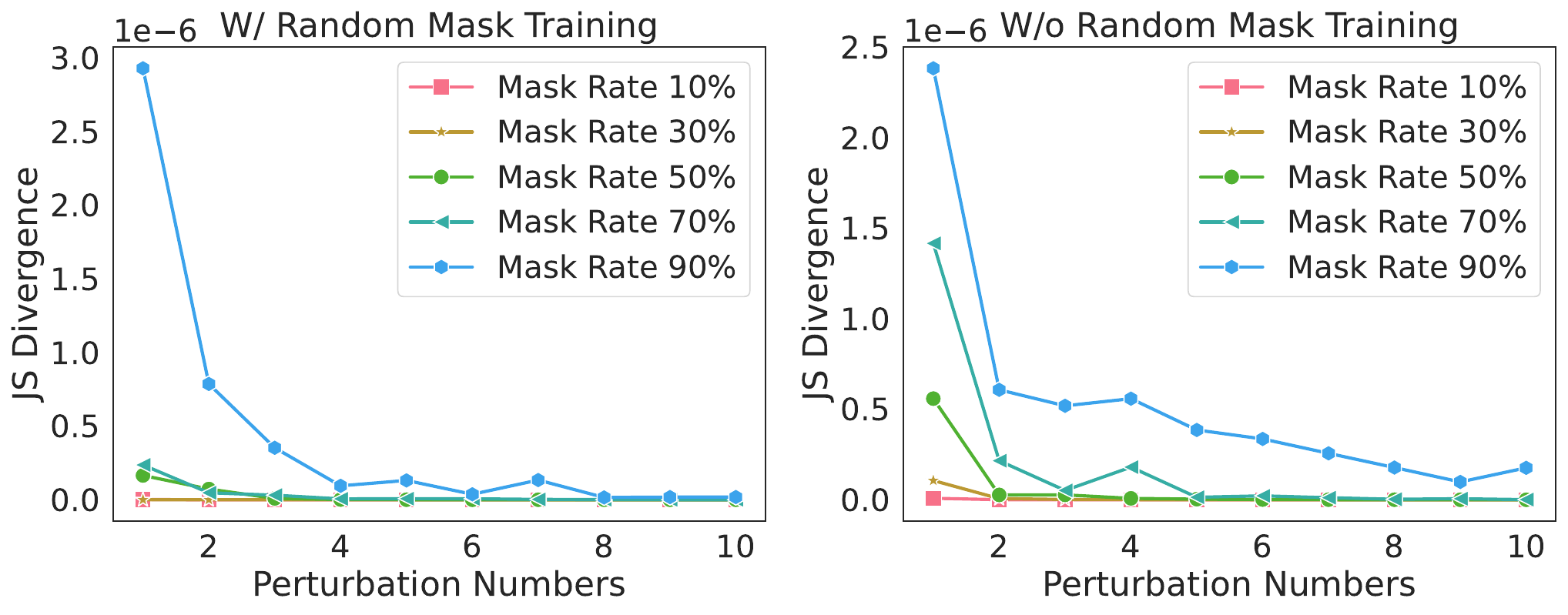}
    \caption{The Jensen-Shannon divergence between $\beta$ and $p_c (x)$ calculated by applying different masking ratios (10\%, 30\%, 50\%, 70\%, and 90\%) on the MSMARCO Passage DEV dataset.}
    \label{fig:beta_estimation}
\end{figure}

As the value of $r$ increases, for any $a$, it becomes increasingly likely to coincide with any randomly sampled $b$, and the value of $\beta$ will approach $g(x')$. To observe the degree of proximity between $\beta$ and $g(x')$, we conducted a simple experiment on the MSMARCO Passage DEV dataset. We randomly selected 100 sampled triplets, set $n_r=500$ and $n_k=1000$, and used Jensen-Shannon divergence to measure the distributional difference between$\beta$ and $g(x')$. As shown in Figure \ref{fig:beta_estimation}, regardless of the masking ratio $\rho$, when the number of perturbed tokens is sufficiently large, all Jensen-Shannon divergences become extremely small (less than $1\times10^{-5}$). Even when the target ranking model was not further fine-tuned on randomly masked ranking data, the Jensen-Shannon divergence condition still held. However, models not fine-tuned on randomly masked ranking data exhibited smoother performance and slightly larger estimation errors under lower masking ratios. Therefore, based on the conclusions from the above validation experiments, we can approximate $g$ using $\beta$ in subsequent experiments, i.e., $\beta\approx g(x')$.

\subsection{Practical Defense Method of Certified Robustness}
Based on the theoretical analysis presented above, we now propose a practically certifiable robust defense method for text ranking models. To ensure that the smoothed ranking model \( \mathrm{g} \) to rank samples accurately and robustly, we first construct samples with a masking rate of \( \rho \) to train the base classifier \( \overline{\mathrm{g}} \). During each iteration of the training process, we sample a mini-batch of data and apply random masking. Subsequently, gradient descent is employed on the masked mini-batch of text to train the classifier \( \overline{\mathrm{g}} \).

\begin{algorithm}[!t] 
        \newcommand\Parameter{\textbf{Parameter: }}
        \newcommand\Functions{\textbf{Functions: }}
        \newcommand\Instructions{\textbf{Instructions: }}
        \caption{Model Prediction and Robustness Certification}  
        \label{alg:algorithm2}  
        \LinesNotNumbered
        \KwIn{query $ \boldsymbol{q}$, document $\boldsymbol{x}$ }
        
        \Parameter{Input text length $T$, Number of reserved words $k$, Number of perturbed words $R$, Number of masked copies for prediction $n$, Number of masked copies for certification $n'$, Number of perturbed samples $n_r$, original ranker $\mathrm{f}$}\\
        \LinesNumbered
        \KwOut{Predicted relevance score $s$ and certifiable robustness radius $r_{rate}$}
        
        \SetKwProg{Proc}{Function}{}{}
        \Proc{\textit{RelPredict}($x, T, k, n, \mathrm{f}$): }{
            $\mathcal{H} \gets $ sample $n$ elements from $U(T, k)$ \\
            $rel\_{list} \gets sample \: n_r \: elements \: from \: \mathcal{U}(T, r)$ \\
            \For{each $h \in \mathcal{H}$}{
                $x_{mask} \gets \mathrm{M}(x, h)$ \\
                $s \gets f(x_{mask})$ \\
                $rel\_{list} \gets $ insert relevance score $s$ \\
        }
        $\overline{\mathrm{g}(x)} \gets $calculate the mean of $rel\_{list}$ \\
        \Return{\( \overline{\mathrm{g}(x)}, rel\_{list} \)}\
        }
        \Proc{\textit{RelRankerCertify}($x, T, k, n, n', n_r, \mathrm{f}$): }{
            $\mathrm{g} \gets RelPredict()$ \\
            \For{$R \gets 0$ to $T$}{
                $\beta \gets BetaEstimator(x, T, k, R, n_r, n', g)$ \\
                $\alpha \gets$ based on equation (9) and value $T, k$ \\
                $\Delta \gets$ based on equation (9) and value $T, R, k$ \\
                \If{$\mathrm{g}(q, \mathcal{M}(x_k, \mathcal{H})) - \mathrm{g}(q, \mathcal{M}(x_{k+1}, \mathcal{H})) \geq \alpha \cdot \beta \cdot \Delta$}{$R \gets R + 1$}
                else break
        }
        \Return{\( r_{rate} = R/T \)}\
        }
        \label{algorithm2}
\end{algorithm}

Specifically, Algorithm \ref{alg:algorithm2} provides a method for estimating $\mathrm{g}(x)$ using Monte Carlo sampling and proving the robustness of $\mathrm{g}$ around $x$. Estimating the predicted value of the smoothed ranking model \( \mathrm{g} \) requires determining the expected value of its predictions. This process is defined in the algorithm pseudo code as the Predict() function, which randomly samples $n$ masked copies and inputs these copies into the original ranking model $\mathrm{f}$ (as part of the classifier $\overline{\mathrm{g}}$), obtaining the average of their predicted values as the prediction of the smoothed model.  In addition, it is also necessary to estimate $\beta$.

As presented in the RelRankerCertify method in Algorithm \ref{alg:algorithm2}, we gradually increase the number of perturbation words R such that $R⁄T$ steadily rises from 0 to 1. We randomly generate $n'$ masked copies of $x$ (where $n'$ is significantly greater than $ n $) and estimate the $\beta$ value through function $BetaEstimator$, calculating the $\Delta$ value using equation (9). Subsequently, we use the smoothed model to rank the query q, comparing the relevance scores of the $k$-th and $k+1$-th documents in the ranking results. According to Proposition 1.1, this process continues until $\mathrm{g}(q, \mathcal{M}(x_k, \mathcal{H})) - \mathrm{g}(q, \mathcal{M}(x_{k+1}, \mathcal{H})) - \alpha \cdot \beta \cdot \Delta < 0$; upon stopping, $R⁄T$ is output as the maximum provable robustness radius of the model $\mathrm{g}$ for x.

\section{Experiment Analysis}
\subsection{Experiment Setup}
\subsubsection{Parameters and Dataset}
The experiments primarily utilize PyTorch and the Transformers library by HuggingFace \cite{wolf2019huggingface} to implement the methods discussed herein and to conduct the corresponding experiments.

The experimental parameters are configured as follows: the maximum concatenated length of queries and candidate documents is set to 256, and the learning rate for model training is selected from 1, 3, 5, 7 × $10^{-6}$. The batch size is set to 256. All models are trained and evaluated on two high-performance servers: Server 1 is equipped with four NVIDIA RTX 24GB 3090 GPUs and 128GB system memory. Server 2 is equipped with four 80GB NVIDIA A100 GPUs and 521GB memory.

Our experiments primarily utilize the MS MARCO Passage dataset and the TREC DL 2019 dataset, both of which are widely used in the field of information retrieval. The MS MARCO Passage dataset, part of the Machine Reading Comprehension (MS MARCO) dataset, is constructed from samples of real user queries on Bing. Initially, this dataset was formed by retrieving the top 10 documents from the Bing search engine and annotating them. The relevance labels in this dataset are determined using a sparse judgment method, assigning labels based on whether the answer to the query appears in the current document, with each document being segmented into passages. The full training set comprises approximately four million query tuples, each consisting of a query, relevant or non-relevant document passages, and their labels. The MS MARCO DEV validation set is primarily used for re-ranking document passages. It includes 6,980 queries, along with the top 1,000 document passages retrieved using BM25 from the MS MARCO corpus. The experiments in this paper employ BM25 retrieval results based on the open-source Anserini library. The average length of passages included in the dataset is 58 words. The TREC DL 2019 dataset is a passage ranking task from the 2019 TREC Deep Learning Track dataset \cite{craswell2020overview}. It shares the same source of candidate documents as the MS MARCO dataset, both derived from Bing search queries and document collections. The difference between it and the MS MARCO dataset lies in the evaluation set organized by TREC DL, which provides 200 distinct new queries, among which 43 queries are furnished with four-level relevance labels. These labels are carefully determined by professional assessors employed by NIST through detailed manual evaluation. For queries with multiple relevant passages, four levels of graded relevance labels are assigned: fully relevant (3), highly relevant (2), relevant (1), and not relevant (0).

\subsubsection{Evaluation Metrics}
In terms of basic evaluation metrics, we first assess the performance of PairLM, following training with different random masking strategy parameters, on the ``clean'' and non-attacked MSMARCO dataset and TREC DL 2019 dataset, in comparison to the BM25 and TK models. This evaluation employs the widely used metrics MRR@10 and NDCG@10 in information retrieval tasks. MRR@10 is used to evaluate the ability of the target model to place the first correct answer as close to the top of the ranking as possible. This is specifically done by calculating the mean of the reciprocal rank across all the queries, where the reciprocal rank is derived from the rank of the first correct answer. A higher MRR@10 value indicates better system performance. On the other hand, NDCG@10 is motivated by the notion that for a search user, highly relevant documents are more valuable compared to marginally relevant ones. It considers the overall ranking performance of the top 10 candidate answers, as opposed to MRR, which only considers the first correct answer. A larger NDCG@10 value signifies better overall performance of the system within the top 10 answers.

To evaluate the robustness of information retrieval models against word substitution attacks, we refer to the Certified Robust Query rate (CRQ) proposed by Wu et al. \cite{wu2022certified}. A query is considered certifiably robust if none of the documents positioned outside the top K are attacked such that they move into the top K positions. Precisely assessing this metric would require enumerating all documents and exponential perturbations, which is computationally prohibitive. As an alternative, CRQ can be evaluated under the condition of randomized smoothing. A higher value of this metric indicates better robustness performance.

Also, to further evaluate the defensive capability of RobustMask, we also employ two metrics during the assessment: Mean Certified Robustness (MCR) and Mean Certified Robustness Radius (MCRR). MCR represents the average certified robustness of RobustMask across all queries, the certified robustness of a query $q$ denotes the maximum $R$ when $|| x' - x ||_0 = R$. MCRR is defined as the average quotient of the certified robustness for each $q$ divided by the length of the target document. Higher values of MCR and MCRR indicate a stronger certified robustness enhancement of RobustMask for retrieval models.

\subsubsection{Baseline}
In the experimental section, we adopt four typical baseline models to analyze the inherent robustness of neural text ranking models and to evaluate the effectiveness of defense methods in enhancing ranking robustness. These four models are highly representative. Our selection includes classic neural ranking architectures (such as BERT ranker) and dense retrieval models (such as BGE). Specifically, we select the Dense Retrieval model BGE (BAAI General Embeddings) as a validation baseline, as it is a widely deployed mainstream model in practice. Although current decoder-only, LLM-based retrieval models demonstrate significant advantages in absolute relevance judgment, their latency and deployment costs mean that, at present, they are not as widely used as mainstream dense retrieval models.

Our choice aligns with current industry trends, where BERT-based architectures still dominate most text ranking scenarios. The main focus of this paper is to improve the robustness of text ranking models, rather than to pursue absolute state-of-the-art performance in retrieval ranking. Meanwhile, the proposed methods can be transferred to other models, which can be further explored in future work. A brief introduction of these four methods is provided below:

(1) BM25 is a classic and widely used term matching algorithm in the field of information retrieval. It is based on the concepts of Term Frequency (TF) and Inverse Document Frequency (IDF), along with normalization of document length, allowing for precise matching between query terms and documents. The strength of BM25 lies in its efficiency and accuracy, especially in the first stage of a two-stage retrieval paradigm (namely the recall phase), where it performs exceptionally well. In our experiments, we utilized parameters specifically fine-tuned for the MS MARCO DEV data and used official experimental results as the benchmark performance for BM25.

(2) TK is a relatively unique information retrieval model, distinguished by its design philosophy, which does not rely on BERT pre-trained models. Instead, it employs a shallow Transformer encoder and the pre-trained word embeddings, diverging from the pre-training paradigm that a significant amount of research currently explores. This design enables TK to maintain model performance while reducing complexity and improving efficiency. The introduction of a Transformer encoder allows TK to handle more complex word sequence relations, while the use of word embeddings provides TK with an initial level of semantic understanding from the early stages of model training. In the experiments of comparing baseline models' performance, we directly adopted the reported performance of the TK model from the work of Hofstätter et al. \cite{hofstatter2020interpretable} as the benchmark performance for TK.

(3) In the previous sections, PairLM has been introduced as a pairwise ranking model utilizing the concatenation approach, which currently achieves the best re-ranking performance. In this method, the query and a pair of candidates are concatenated using [SEP] and [CLS] tokens, respectively. Essentially, a shared parameter LM encoder is trained on these two concatenated inputs. The relevance scores are additionally obtained via a trainable linear layer, allowing for the acquisition of both positive and negative relevance scores. During the training phase, gradients are calculated through back propagation, and iterative optimization is conducted using the Adam optimizer. In the inference phase, the relevance score for a document corresponding to a query is derived by selecting the final dimensional output of the relevance scorer.

(4) BGE is a previous state-of-the-art embedding model. The BGE model is initially pre-trained on large-scale text corpora using MAE-style learning framework, followed by contrastive learning optimization and task-specific fine-tuning to enhance its semantic representation capabilities. This hierarchical training paradigm establishes BGE as a high-performance embedding model, achieving state-of-the-art results in semantic similarity tasks. Consequently, BGE remains extensively employed as a core retrieval component in industrial and research applications, including search engines, open-domain question answering systems, and retrieval-augmented generation pipelines for LLMs. In the experiment, we adopt the BGE-M3 model from FlagEmbedding as one of the baseline models.

\subsection{Analysis and Discussion}
To validate the effectiveness of RobustMask for robustness enhancement in information retrieval models, several research questions are addressed in detail: (1) How does the performance of the ranking model with random masking smoothing compare to the original model? (2) How is the Top-K certified robustness at different mask ratios? (3) How does the certifiable robustness enhancement method, RobustMask, perform in defending against actual information manipulation attacks compared to other methods?

\subsubsection{RQ1: How does the performance of the ranking model with random masking smoothing compare to the original model?}

\begin{table}
  \caption{The comparison of conventional ranking performance between the original ranking model and various smoothed ranking models on the MS-MARCO Passage DEV dataset and the TREC DL 2019 dataset.}
  \label{tab:mask_train}
  \centering
  \resizebox{0.49\textwidth}{!}{
  \begin{tabular}{ccccc}
    \toprule
    \multirow{2}{*}{Method} & \multicolumn{2}{c}{MSMARCO DEV} & \multicolumn{2}{c}{TREC DL 2019}  \\
      & MRR@10 & NDCG@10 & MRR@10 & NDCG@10 \\
     \midrule
      BM25 & 18.7 & 23.4 & 68.5 & 48.7 \\
      TK & 33.1 & 38.4 & 75.1 & 65.2 \\
      BERT-Base & 35.2 & 41.5 & 87.1 & 71.0 \\
      BGE & - & - & 85.9 & 67.9\\ 
      BERT-Base+PGD & - & - & 82.6 & 66.8 \\
      PairLM (BERT-Base) & 34.3 & 40.4 & 85.7 & 71.2 \\
      +CertDR w/o Data Augments & 19.5 & 23.5 & 66.7 & 54.2 \\
      +CertDR & 31.9 & 36.9 & 77.4 & 66.1 \\
      +Random Mask 30\% Ranker & 34.0 & 40.2 & 84.5 & 71.7 \\
      +Random Mask 60\% Ranker & 34.1 & 40.3 & 83.6 & 71.5 \\
    \bottomrule
  \end{tabular}
  }
\end{table}

The performance metrics for BM25, TK, BGE and BERT-Base presented in Table 1 are derived from the replication of results from corresponding open-source projects, or reported outcomes in existing literature, or our own experiment results. The original performance of PairLM (BERT-Base) was obtained by training the PairLM model on the official MSMARCO triplet training set. The performance degradation of the original PairLM (BERT-Base) compared to BERT-Base can be attributed to two primary factors: overfitting to the triplet training set and differences in the optimization techniques and number of GPU machines utilized during the training process of the BERT-Base ranking model. Furthermore, we also present the performance of BERT-Base following the application of adversarial training via Projected Gradient Descent (PGD) with continuous embedding perturbation.

\begin{figure}[!t]
\centering
\includegraphics[width=0.98\linewidth]{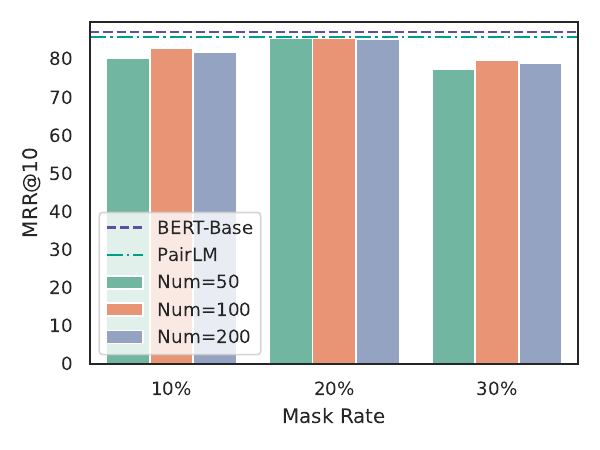}
\caption{The experimental results comparing the performance of smoothed ranking models on the clean TREC DL 2019 dataset under different random masking rates and varying numbers of ensemble samples.}
\label{fig_mrr}
\end{figure}

The focus of this research is not on optimizing performance on clean datasets where performance is considered acceptable, but rather on examining the efficacy of robustness enhancement methods. The focus of this research is not on optimizing performance on clean datasets where fundamental ranking performance needs only be acceptable, but rather on examining the efficacy of robustness enhancement methods. Specifically, this investigation aims to assess the improvements in the defense and resilience of the target ranking models against information manipulation attacks.

In typical datasets, these [MASK] tokens are not usually present. Therefore, evaluating smoothed models using clean data without [MASK] tokens may have a slight impact on the results. The discrepancy between the data used during evaluation and that employed in training may lead to slight variations in the model performance. As demonstrated in Table 1, the model trained with a 0.6 masking ratio, when compared to the PairLM trained on the normal dataset, exhibited a decrease of 0.2 percentage points in MRR@10 and a decrease of 0.1 percentage points in NDCG@10 on the MSMARCO DEV dataset. On the TREC DL 2019 dataset, the MRR@10 decreased by 1.2 percentage points, while NDCG@10 increased by 0.3 percentage points. The performance degradation is primarily due to inconsistencies between training and evaluation processes. Since normal datasets do not contain Mask tokens, this can slightly affect the model's performance on clean data. However, overall, our method still achieves performance levels comparable to the original target model, PairLM (BERT-Base). Compared with CertDR \cite{wu2022certified}, a certified robustness method based on word substitution for constructing smoothed models, RobustMask exhibits distinct comparative advantages. In the case of CertDR and its data augmentation-based variant, despite being based on a strong assumption of knowing the substitution attack vocabulary used by the information manipulation attacker, this word substitution approach still significantly impairs the performance of the target model. It requires more normal data to enhance its semantic understanding capabilities in conventional ranking tasks. This result also demonstrates that our RobustMask method minimally diminishes the model's abilities in normal scenarios, maintaining robust performance even when faced with the challenge of training and evaluation inconsistencies.

\begin{figure*}[!t]
\centering
\includegraphics[width=1\linewidth]{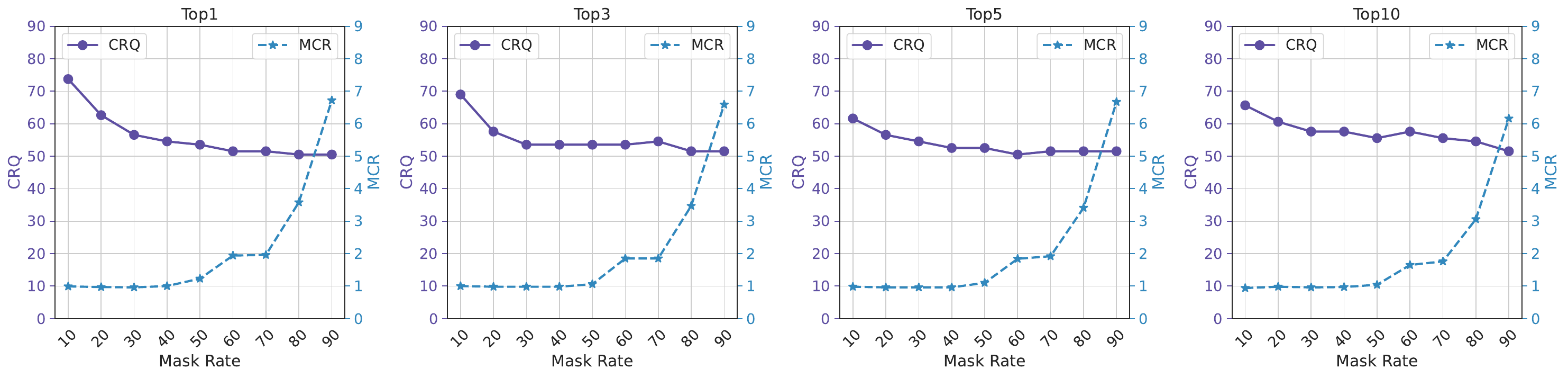}
\caption{Top-K CRQ and the corresponding MCR under different Mask rates on the MSMARCO dataset.}
\label{fig_crq2mcr}
\end{figure*}

\begin{figure*}[!t]
\centering
\includegraphics[width=1\linewidth]{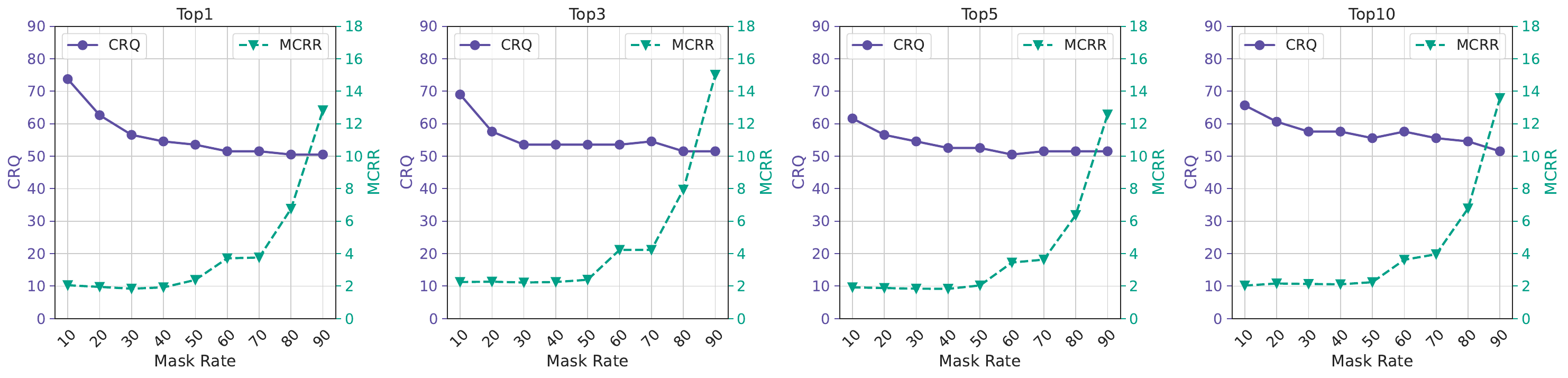}
\caption{Top-K CRQ and the corresponding MCRR under different Mask rates on the MSMARCO dataset.}
\label{fig_crq2mcrr}
\end{figure*}

\begin{figure*}[t]
  \centering
  \includegraphics[width=\linewidth]{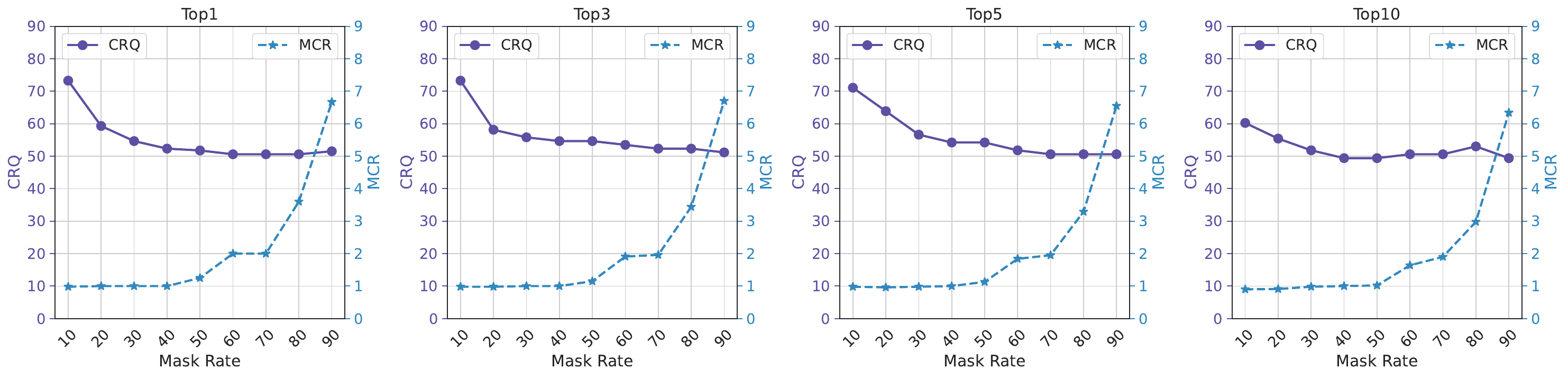}
  \caption{Top-K CRQ and the corresponding MCR under different mask rates on the TREC DL 2019 dataset.}
  \label{fig_crq2mcr_dl}
\end{figure*}

\begin{figure*}[t]
  \centering
  \includegraphics[width=\linewidth]{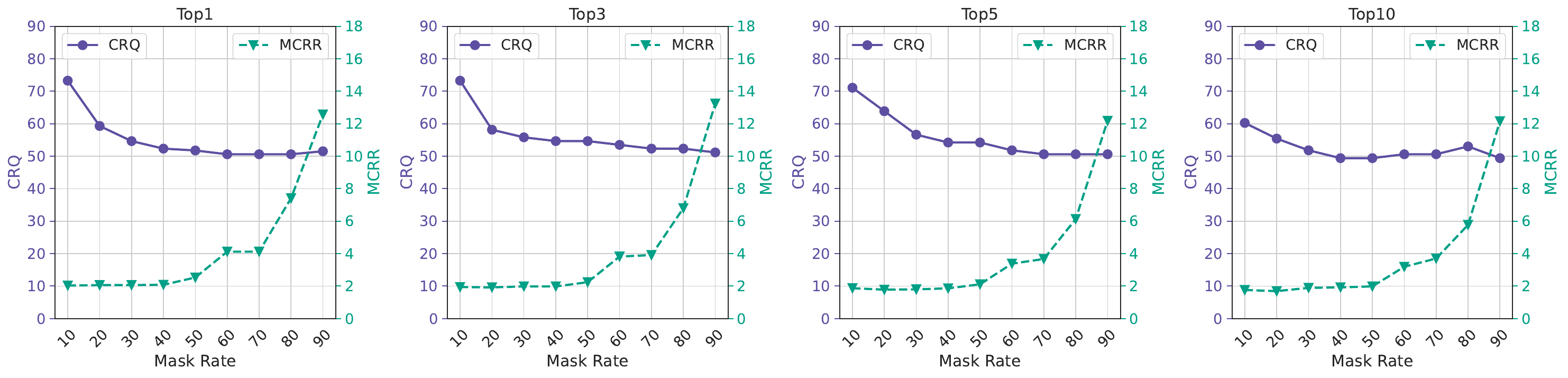}
  \caption{Top-K CRQ and the corresponding MCRR under different mask rates on the TREC DL 2019 dataset.}
  \label{fig_crq2mcrr_dl}
\end{figure*}

Due to the large number of queries in the MSMARCO DEV data, which results in extended testing times when combined with the ensemble numbers, the experimental comparison of the performance of smoothed ranking models under different random masking rates and varying ensemble sample sizes was conducted solely on the TREC DL 2019 dataset.  
Figure \ref{fig_mrr} reveals that the performance of the BERT-Base smoothed ranking model shows a minor decline compared to the non-random masking integrated PairLM (BERT-Base) in most cases. Specifically, with a masking rate of 20\%, the MRR@10 decreases less significantly, indicating that random smoothing has minimal impact on top-ranked documents. In addition, since the employed neural ranking model is lightweight, RobustMask has low overhead with short runtime and its memory consumption and time cost can be further optimized by GPU-based parallel acceleration. 

Figure \ref{fig_mrr} evaluates three sample sizes for random masking integration, namely {Num=50, Num=100, Num=200}. The results indicate that MRR@10 varies only slightly across these settings, suggesting that once a statistically representative number of samples is integrated, the overall ranking performance remains largely stable.

\subsubsection{RQ2: How is the Top-K certified robustness at different mask ratios?}
We conducted experiments on the MSMARCO DEV and TREC DL 2019 datasets to evaluate the Top-K (K={1, 3, 5, 10}) Certified Robust Query rate (CRQ) under different random masking rates. Additionally, we measured the corresponding Mean Certified Robustness (MCR) and Mean Certified Robustness Radius (MCRR). The Top-K Certifiable Robustness for a given query \(Q\) refers to the maximum radius \(R\) that the smoothed relevance judgment model \(g\) can provide for the triplet \(x\) composed of the query, the K-th document, and the K+1-th document. Here, \(x^\prime\) represents an adversarial example, and for any \(||x - x^\prime||_0 \leq R\), the model can produce a correct relative relevance judgment.

From Figures \ref{fig_crq2mcr} and \ref{fig_crq2mcrr}, it can be observed that with the increase in the mask ratio, there is a decrease in the Certified Robust Query (CRQ) ratio on the MSMARCO dataset. This indicates that the increase in masks complicates the difficulty of relevance determination, leading to a reduction in statistically significant and reliable relevance assessments. Consequently, the model is more likely to refuse predictions, resulting in a higher frequency of ``unverifiable'' judgments. However, concurrently, as the certifiable robust perturbation radius increases, more adversarial modifications are necessary to successfully attack a ranking model, thereby incurring greater costs. More results on TREC DL 2019 dataset are depicted Figure \ref{fig_crq2mcr_dl} and Figure \ref{fig_crq2mcrr_dl} at appendix.

Figures \ref{fig_crq2mcr_dl} and \ref{fig_crq2mcrr_dl} present the corresponding results on the TREC DL 2019 dataset. Overall, the observed trends are consistent with those on MSMARCO. As the mask ratio increases, the Certified Robust Query (CRQ) ratio declines, leading to more frequent ``unverifiable'' judgments, while the certifiable robust perturbation radius expands, implying higher adversarial costs. The main difference lies in the specific numerical values, which vary due to the characteristics of the TREC DL 2019 dataset.

\subsubsection{RQ3: How does the certifiable robustness enhancement method, RobustMask, perform in defending against actual information manipulation attacks compared to other methods?}
For RQ3, we focus on BERT and BGE as the primary target ranking models to examine different robustness enhancement baseline methods and ranking attack methods: (1) The original target ranking model (BERT and BGE) serves mainly as a baseline to assess the ability of the original model to withstand information manipulation attacks in the absence of any defense methods; (2) Models (BERT only) enhanced by robustness enhancement methods based on adversarial training, such as PGD adversarial training (PGD-ADV) and adversarial training based on PAT samples (PAT-ADV), which correspond to continuous perturbation adversarial training and discrete perturbation adversarial training, respectively; (3) The proposed certifiable robustness enhancement method, RobustMask, based on random masking for text ranking models, primarily combines the innate characteristics of pre-trained models to mitigate the effects of various adversarial attack perturbations through robust random mask smoothing.

The actual information manipulation attack methods primarily include keyword stuffing (Query+), adversarial semantic collision (Collision\_nat), Pairwise Anchor Trigger(PAT) attack methods, PRADA, among other attack methodologies. It is important to note that for information manipulation attacks based on trigger insertion, such as keyword stuffing and adversarial semantic collision, adjustments were made in the experiments: the generated triggers were inserted at the beginning of candidate documents, with equivalent lengths of text being truncated from the end to achieve a ``word replacement'' attack transformation. The hyperparameters such as trigger length and search space follow the settings of the experiments discussed in earlier sections, with adversarial perturbation content not exceeding 5\%. The experimental results on TREC DL 2019 are shown in Tables 3 and 4.

\begin{table}[!t]
  \caption{The results of the Top-10 defense success rate (\%) against empirical attacks using robustness enhancement methods on the TREC DL 2019 dataset.}
  \label{tab:top10asr}
  \centering
  \resizebox{0.49\textwidth}{!}{
  \begin{tabular}{ccccc}
    \toprule
    Method & Query+ & Collision\_nat & PAT & PRADA \\
     \midrule
      BERT-base & 92.9 & 51.2 & 29.3 & 78.6 \\
      BGE & 97.6 & 68.3 & 31.7 & 35.7 \\
      PGD-ADV & 73.8 & 75.6 & 31.7 & 38.1 \\
      PAT-ADV & 57.1 & 43.9 & 26.8 & 31.0 \\
      RobustMask-30\% (BERT) & 64.3 & 31.7 & 12.2 & 38.1 \\
      RobustMask-60\% (BERT) & 33.3 & 22.0 & 14.6 & 33.3 \\
      RobustMask-90\% (BERT) & 11.9 & 9.8 & 9.8 & 23.8 \\
      RobustMask-30\% (BGE) & 83.3 & 36.9 & 4.87 & 21.9 \\
      RobustMask-60\% (BGE) & 14.3 & 4.9 & 2.4 & 2.4 \\
      RobustMask-90\% (BGE) & 0.0 & 0.0 & 0.0 & 0.0 \\
    \bottomrule
  \end{tabular}
  }
\end{table}

It can be observed that if no defensive measures are taken to enhance the robustness of the base model(BERT-Base and BGE), the success rate of information manipulation adversarial attacks is remarkably high. For instance, on the TREC DL 2019 dataset, the attack success rate of Query+ against BERT-base reached 90\%. This underscores the necessity of proposing appropriate defense methods. Although empirically-based adversarial training defense methods reduce the success rate of empirical attacks to some extent, their performance does not match that of RobustMask. Therefore, merely increasing training documents and adversarial samples is not a robust defensive strategy against adversarial information manipulation attacks in information retrieval tasks. Future robustness enhancement methods in the field of information retrieval should explore empirical defense strategies that are more suitable for retrieval ranking scenarios. However, empirical defense methods cannot provide strict certifiable robustness guarantees, and their performance may largely depend on the dataset and the specific attack method. For example, for keyword stuffing (Query+), a commonly used attack method in search engine optimization, general end-to-end defense ranking models find it challenging to defend effectively. The defense is typically possible only by deploying empirically-based information manipulation attack anomaly detectors explored previously. In contrast, RobustMask enables end-to-end defense against keyword stuffing for target ranking models. Moreover, RobustMask can also effectively defend against other attack methods based on gradients and important word replacements. Particularly, for PAT attacks, which are less effective at inverting adjacent positional relevance, the attack success rate could be low due to incorporating camouflage and contextual consistency mechanisms, sacrificing some attack success rate. Our proposed RobustMask can further reduce the attack success rate of PAT, building on its already low success rate. This indicates that RobustMask can theoretically certify the robustness of ranking models and practically enhance their robustness.

\begin{table}[!t]
  \caption{The results of the Top-20 defense success rate (\%) against empirical attacks using robustness enhancement methods on the TREC DL 2019 dataset.}
  \label{tab:top20asr}
  \centering
  \resizebox{0.49\textwidth}{!}{
  \begin{tabular}{ccccc}
    \toprule
    Method & Query+ & Collision\_nat & PAT & PRADA \\
     \midrule
      BERT-base & 95.2 & 70.7 & 24.4 & 85.7 \\
      BGE & 97.6 & 53.6 &	24.4 & 30.9 \\
      PGD-ADV & 88.1 & 68.3 & 22.0 & 64.3 \\
      PAT-ADV & 71.4 & 51.2 & 17.7 & 50.0 \\
      RobustMask-30\% (BERT) & 81.0 & 43.9 & 9.8 & 28.6 \\
      RobustMask-60\% (BERT) & 57.1 & 41.5 & 26.8 & 23.8 \\
      RobustMask-90\% (BERT) & 38.1 & 26.8 & 24.4 & 21.4 \\
      RobustMask-30\% (BGE) & 83.3 & 26.8 & 7.3 & 7.3 \\
      RobustMask-60\% (BGE) & 33.3 & 7.3 & 2.4 & 0.0 \\
      RobustMask-90\% (BGE) & 2.4 & 0.0 & 0.0 & 0.0 \\
    \bottomrule
  \end{tabular}
  }
\end{table}

Furthermore, from the tables of the two experimental results mentioned above, it can be observed that the defense performance of RobustMask is relatively sensitive to the probability of random mask smoothing. The optimal masking probability varies among different methods, primarily because the current certifiable methods have a small robustness radius, and different attack methods result in varying adversarial modifications. In practical application, some costs need to be incurred to adjust this parameter. Future research could explore defense methods with a larger certifiable robustness radius, aiming to reduce method specificity and thereby enhance the robustness of target models more efficiently.

\section{Conclusion}
In this paper, we introduced RobustMask, a novel defense approach designed to enhance and certify the robustness of neural text ranking models against various adversarial perturbations at the character, word, and phrase levels. Leveraging the intrinsic masking prediction capability of pretrained language models with a strategically randomized smoothing mechanism, RobustMask significantly mitigates potential adversarial manipulation, effectively shielding retrieval systems from harm. We rigorously established theoretically provable certified robustness guarantees by integrating pairwise ranking comparisons and probabilistic statistical methods, enabling formal verification of top-K ranking stability under bounded adversarial perturbations. Empirical evaluations also demonstrated its effectiveness, maintaining certified robustness for the top-10 and top-20 ranked candidates. These results significantly surpass the capabilities of existing methods, highlighting the practical usability and substantial advantage of our method in real-world retrieval and RAG scenarios. Our study advances a critical step toward ensuring the secure deployment of neural ranking models in adversarial environments, paving the way for trustworthy and stable information retrieval systems and their downstream applications. In the future, we will explore further refinement of masking strategies and adaptive certification schemes, enhancing certified robustness in more target models, dynamic scenarios, and across diverse adversarial environments.

\bibliographystyle{plain}
\bibliography{ref}


\appendix

\end{document}